# Match predictions in soccer: Machine learning vs. Poisson approaches

Mirko Fischer[a], Andreas Heuer[a]

[a]Institute of Physical Chemistry, University of Münster,
Corrensstrasse 28/30, 48149 Münster

Correspondence: mirko.fischer@uni-muenster.de, andheuer@uni-muenster.de

**Summary**

Predicting the results of soccer matches is of great interest. This is not only due to the popularity of the sport and the joy of private "betting rounds", but also due to the large sports betting market. Where previously expert knowledge and intuition were used, today there are models that analyze large amounts of data and make predictions based on them. In addition to Poisson models, approaches that belong to the machine learning (ML) category are increasingly being used. These include, for example, neural network or random forest models, which are compared in this article with each other as well as with Poisson models with regard to single-match prediction. In each case, the match results of a season are used as the data basis. The analysis is carried out for 5 European top leagues. A statistical analysis shows that the performance levels of the teams do not change systematically during a season. In order to characterize performance levels as accurately as possible, all match results, except from the match to be predicted, can be used as features with equal weighting. It can be seen that both, the exact choice of features and the choice of model, have only a minor influence on the prediction quality. Possible improvements in match prediction are discussed.

## 1 Introduction and motivation

### 1.1 Relevance of the match predictions

Since the 1960s, attempts have been made to systematically predict the results of soccer matches using statistical models (Dubitzky 2019), while previous predictions were essentially based on "expert knowledge". One of the main reasons for the high level of interest in such models and predictions is the growing international sports betting market, which is worth several billion US dollars per year (Etuk 2022).

A match prediction essentially consists of two steps. First, the most meaningful information possible must be collected about the two teams. Matches are now analyzed and quantified in detail, especially using automated techniques (Goes 2020). Teams can use this data to make important tactical decisions or identify suitable players for potential future transfers. In the context of match prediction, the performance strengths of the teams in particular can be estimated. Due to the limited information available before the match, the assessment of the performance of the two teams will always be subject to a certain degree of inaccuracy, but this can at least be minimized by optimizing the data used.

In the second step, a match prediction model is required, which is "fed" with the available data. This shows that the result is not completely predictable. Reep and Benjamin (Reep 1968) already recognized that chance plays a significant role in the outcome of a game. For example, the teams with the higher market value do not always win - this is a major part of the appeal of soccer. Therefore, the actual prediction consists of determining probabilities for all possible outcomes. There are various fundamental reasons why exact predictions are not possible (Heuer 2012, Heuer 2014).

(1) Relatively few goals are scored (on average between 2 and 3 goals in European professional leagues), so that the match result is generally a consequence of relatively few relevant actions and referee decisions, for example, have a major influence. This effect would disappear if (hypothetically) matches lasted much longer with the same level of performance.
(2) There are irreversible match-immanent effects (red cards, in-game injuries) that are unpredictable but can have a significant impact on the final result.
(3) There are match day-specific effects such as the prior absence of players due to injury or specific match tactics whose impact on the expected match result can hardly be quantified. All of these reasons can be summarized as random effects.

We can speak of a *perfect* tip if both steps are optimal, i.e. the estimation of the previous performance level and the model are optimal. Differences in the quality of the two points explain, for example, the slight differences that were observed in the "Soccer Prediction Challenge" (see below).

**1.2 Machine learning in match prediction**

Poisson models imply that the random effects relating to the goals scored can essentially be described according to Poisson statistics. In contrast, data-driven approaches, used here synonymously with machine learning (ML), do not make these assumptions. Another advantage is the simple integration of the ever-growing volumes of data in professional soccer. The first ML models were developed in the early 2010s. Kumar (Kumar 2013) compared various ML models for predicting soccer results. In the following years, further models were presented and tested. The aim is often to predict the outcome of a match based on previously known data. However, some models also use data collected during a match (e.g. strength of tackles, ball possession, shots on goal, etc.) to increase the predictive power.

Here is a brief overview of some models:

- Poisson (Goddard 2005, Koopman 2015, Angelini 2017)
- Bayesian Networks (Joseph 2006, Constantinou 2012, Constantinou 2019)
- Neural networks (Mendes-Neves 2020, Jain 2021)
- Random Forest (Stübinger 2020)
- K-Nearest-Neighbor (Esme 2018, Berrar 2019)
- Boosting (XGBoost, CatBoost) (Hubacek, 2019, Ren 2022, Yeung 2023)

In 2018, Dubnitzky et. al launched the "Soccer Prediction Challenge", in which various teams were able to submit models that were supposed to predict the outcome of future matches based on previous results from 52 different leagues. (Dubnitzky 2018) Only information on the times and results of previous matches could be used as input for the models. When comparing all the models described in the literature (Berrar 2019, Hubaceck 2019, Constantinou 2019, Tsokos 2019), it is noticeable that they show only slight differences in predictive power (RPS score 0.2054 to 0.2087, accuracy: 51.46 to 53.88), although different features were used in some cases. A similar observation is described in the review article by Hubácek (Hubácek 2022). The challenge was repeated in a similar form in 2023. The results can be viewed at https://sites.google.com/view/2023soccerpredictionchallenge/home. The aim of our article is to analyze the second step of the match prediction, i.e. the model selection, whereby different features are also considered with the same data set.

**2 Data set**

In the following, a data set is considered, which consists of over 14,000 match results from the Bundesliga (Germany), Premier League (England), La Liga (Spain), Serie A (Italy) and Ligue 1 (France).

The data was collected over a period of 8 seasons (2014/2015 - 2021/2022). The 2019/2020 season was canceled in France and was therefore not included in the analysis.

**3 Team performance over the course of the season**

The average goal difference that a specific team would score against an average team after averaging over all random effects can be chosen as a useful measure of the performance strength of a team within a league (Heuer 2012). To find out whether the performance of a team changes within a season, the league-averaged autocorrelation function of the goal difference $\Delta G$

$$corr_{\Delta G}(\Delta n) = \langle \Delta G(n) \Delta G(n + \Delta n) \rangle_{t, \text{teams}, \text{leagues}}$$

$$= \frac{1}{\#\text{terms}} \sum_{\text{leagues}} \sum_{m=1}^{N_{\text{teams}}} \sum_{n=1}^{N_{\text{match days}} - \Delta n} \Delta G_m(n) \cdot \Delta G_m(n + \Delta n)$$

can be used, whereby $N_{\text{teams}}$ describes the number of teams and $N_{\text{match days}}$ describes the number of match days within a league. The argument $\Delta n$ describes the time difference and thus records whether and to what extent the performance level changes after matches. $\Delta G_m(n)$ describes the goal difference from the point of view of team m on match day $n$. Some technical aspects are still important: (1) Terms are omitted from the sum if it is the same opponent of team m on match days $n$ and $n + \Delta n$ are involved. (2) $\Delta G_m(n)$ is corrected by the home advantage available for the corresponding season. (3) Especially in the Corona years, the sequence of games is often different than originally planned. Therefore, the temporal argument (e.g. n=10) generally describes the 10th game of team m, which does not necessarily have to take place on the 10th matchday.

The autocorrelation function has a finite positive value if the performance of a team is still positively correlated after a number of match days. If both are completely uncorrelated after a time $\Delta n$ the autocorrelation function disappears.

As can be seen in Fig. 1a, the autocorrelation function shows no significant dependence on $\Delta n$ but fluctuates around a constant value. This results in the surprising observation that the performance of a team does not change during a season in the context of statistical errors. This observation is consistent with preliminary earlier studies (Heuer 2009, Heuer 2012). It should be noted that this does not rule out fluctuations in performance from game to game - but without any systematic time dependency. On the other hand, systematic changes in performance levels do occur when comparing successive seasons (Heuer 2009).

**4 Single match predictions**

**4.1 Prediction of match outcome, goal difference and exact match result**

There are different ways of predicting the outcome of a game. In the simplest variant, only the result in the form win/draw/loss is considered. The prediction of the goal difference is somewhat more specific. In the most informative variant, the exact match result is predicted, as already mentioned, by specifying the probabilities for all possible outcomes. To predict the exact match result using data-driven models, more data is naturally required than when restricting to goal differences, as the number of possible exact match results is significantly larger than the number of possible goal differences. In this paper, we focus on the prediction of goal differences and will also discuss the prediction of total number of goals for comparison.

## 4.2 Neglecting the chronological order

When predicting the results of the last game, all previous games of the teams are available. Since (see above) the performance of a team does not change systematically during a season (within the statistical inaccuracies), the order of the matches is not statistically relevant. Consistency means that a stronger consideration of the last matches does not significantly improve the prediction (Heuer 2012). For our model investigations, we use the information about all other matches of the involved teams, regardless of whether the match is before or after the match to be predicted. Of course, the results can be directly generalized to include only the previous results, see also (Berrar 2018, Hubacek 2019).

## 4.3 Feature engineering

Before training a model, the available data (time of the game, teams involved and final result) must be prepared in such a way that it can be processed automatically.

In this specific case, very simple features are used for illustration. For each match $i$ between teams A (home team) and B, the average goals scored and conceded by the two teams ($x^i_{+,A}$, $x^i_{-,A}$, $x^i_{+,B}$, $x^i_{-,B}$) during the other matches of the season are calculated, e.g.

$$x^i_{+,A} = \frac{1}{N_{\text{match days}} - 1} \sum_{\substack{j=1 \\ j \neq i}}^{N_{\text{match days}}} g^j_{+,A}$$

$$x^i_{-,A} = \frac{1}{N_{\text{match days}} - 1} \sum_{\substack{j=1 \\ j \neq i}}^{N_{\text{match days}}} g^j_{-,A}$$

where here $g^j_{+,A}$ and $g^j_{-,A}$ are the goals scored and conceded by team A in match j. When averaging, it does not matter whether the matches are before or after match i in terms of time.

The features obtained in this way can be further combined. The difference between $x^i_{+,A}$ and $x^i_{-,A}$ results in a new feature $x^i_{\Delta G,A} = x^i_{+,A} - x^i_{-,A}$. Equivalent also $x^i_{\Delta G,AB} = x^i_{\Delta G,A} - x^i_{\Delta G,B}$ can be obtained. In the hypothetical limiting case of an infinitely long season, i.e. the determination of random effects, $x^i_{\Delta G,AB}$ corresponds to the expected goal difference of the match $i$ when averaged over a large number of realizations of the match (Heuer 2012). Similarly, the feature associated with the total number of goals of a team $x^i_{\Sigma G,A} = x^i_{+,A} + x^i_{-,A}$ can be defined analogously. As shown in (Heuer 2012), in the hypothetical limiting case outlined above, the variable $x^i_{\Sigma G,AB} = x^i_{\Sigma G,A} + x^i_{\Sigma G,B} - \langle G \rangle$ is a consistent estimate of the expected goals in the match i under consideration, where $\langle G \rangle$ is the mean number of goals during the season under consideration. In the approach chosen here, the same amount of information is available for each game. Here, all teams are considered equally, which increases the transferability of the model to data from other leagues and seasons. Furthermore, the introduction of a team index does not lead to any improvement in the prediction.

For some applications, the goals are adjusted for home advantage (obtained as the average of all games except game *i*). This removes the effect of how often a team has played at home. This analysis is based on the finding that home advantage is not team-specific (but certainly season-specific) (Heuer 2010, Heuer 2012). The corresponding variables are abbreviated with the symbol $ẋ$ (place of x). The home advantage indicates how many more goals the home team scores on average than the away team.

At the transition from $x^i_{+,A}$ and $x^i_{-,A}$ to $x^i_{\Delta G,A}$ the information is reduced. On the one hand, this makes modeling easier and requires fewer adjustable parameters; on the other hand, it is associated with a loss of information, which could potentially reduce the prediction quality in the borderline case of many data. The actual modeling can now be carried out for different selections of features. By comparing the respective prediction qualities, the extent to which the individual features influence the match result and are therefore relevant for a soccer match can be derived directly.

Many other features are discussed in the literature, which often explicitly include time dependency. Some of them are listed here:

- Historical form
- Importance of a game
- League
- Home advantage
- Page rank (Hubacek 2019)
- ELO rating (Robberechts 2019)
- Pi rating (Constantinou 2012, Hubacek 2019)
- Berrar rating → Characterization of team performance through dynamic ratings that are adjusted after each game (Berrar 2018)

The paper by Yeung (Yeung 2023) provides a good overview of the features used in literature.

**4.4 Models for match prediction**

Three different models are compared for the match prediction: Neural networks (1 intermediate layer with 8 neurons, activation function: logistic, solver: ADAM), Random Forest (100 decision trees, maximum depth: 4, min. 32 data points for a split) and the statistical Poisson model.

The two ML models are used for classification in this context. This means that each goal result is regarded as a class. For each class, the models predict the probability with which it corresponds to the true class based on the input features.

To train the models, the error in the prediction is minimized. More precisely, a so-called loss function is minimized in which the prediction and the true value are compared with each other. In our example, we use cross entropy, which compares the predicted probabilities with the true probabilities (here 1 for the true result and 0 otherwise) for each class.

The cross entropy is given as

$$L(p,q) = -\sum_i p_i \log q_i$$

where $p_i$ is the true and $q_i$ is the predicted probability for each class $i$. If $j$ corresponds to the actual match result, then $L(p,q) = -\log q_j$ is chosen.

In contrast to ML models, which have only been increasingly used for match prediction in recent years, Poisson models have been used successfully for some time (Goddard 2005, Heuer 2010, Koopman 2015, Angelini 2017). For the prediction of game k in the Poisson model, first the home-advantage adjusted variables $\varkappa^i_{g,A}$ and $\varkappa^i_{k,A}$ for team A and analogously for team B for each game $i \neq k$ are calculated. Game k is excluded from the averaging process. For all games *i* (between teams C and D), the linear superposition $a_{\Delta,\text{home}} \varkappa^i_{\Delta G,CD} + a_{\Sigma,\text{home}} \varkappa^i_{\Sigma G,CD} + a_{0,\text{home}}$ is compared with the actual number of goals scored by team C and, after taking all games i into account, the 3 coefficients

$a_{\Delta,\text{home}}$, $a_{\Sigma,\text{home}}$ and $a_{0,\text{home}}$ are determined. The same procedure is used for the goals scored by the away teams. The expected values of the goals of team A are then determined for match $k$ (based on $x^k_{\Delta G,AB}$, $x^k_{\Sigma G,AB}$) and similarly for team B. Then using the Poisson distribution with the expectation value $\lambda$

$$P_\lambda(G) = \frac{\lambda^G}{G!} e^{-\lambda}$$

the probability distribution of the goals $G$ scored by teams A and B is calculated. The simple assumption that the goals of the two teams are uncorrelated is used here. This gives a probability for each result and therefore in particular for each goal difference and the total number of goals.

**4.5 Metrics**

To compare the different models, we determine two different metrics. The cross entropy is already known from the loss function. In addition, we determine the RPS score (ranked probability score), which takes into account the probability distribution of the prediction and the order of the classes:

$$RPS = \frac{1}{n_c - 1} \sum_{i=1}^{n_c} \left( \sum_{j=1}^{i} q_j - \sum_{j=1}^{i} p_j \right)^2$$

$n_c$ indicates the number of classes, $p_j$ the true and $q_j$ the predicted probability for each class $j$.

**4.6 Comparison of models and features**

The cross-validation method is used to compare the models based on the metrics. It allows a statistically more accurate determination of the metrics including standard errors compared to a simple calculation of the metrics using a single test set.

To do this, the data set is divided into $N_{\text{match days}}$ subsets, each representing one match day. Based on this, models are now trained on $N_{\text{match days}} - 1$ match days (training set) and the metrics for a match day (test set) are calculated. This is repeated so that each match day acts as a test set once. The final metrics are averaged over all match days and compared including standard errors in Table 1 for different models and features. Five randomly selected seasons are not included in the calculation of these metrics and are instead used to optimize the hyperparameters. The results are shown in Table 1.

For the prediction of goal differences (-10 to +10), all models are fundamentally better than a simple base model in which the same probabilities $p(\Delta G)$ are used for each game. $p(\Delta G)$ is the normalized frequency with which $\Delta G$ occurs in the data set.

It is also observed that the use of several features does not lead to any significant improvement in the prediction. In general, the neural network model provides slightly better results than the random forest and performs similarly compared to the Poisson model.

Instead of a single model for all leagues, several models can also be trained for each league individually. The metrics differ slightly depending on the league (the more balanced the league, the more difficult the prediction) but are generally worse due to the smaller training set and have a higher standard error (cross entropy of 1.90±0.03 to 2.03±0.03) compared to the model trained on all leagues.

Similar results can be observed for the prediction of the total number of goals (taking into account values from 0 to 16 goals). However, the models are only slightly better in their predictions than the base model defined analogously to above. This observation implies that the total number of goals in a match is less team-specific than the goal difference. Betting on the number of goals in a match is therefore similar in nature to simply rolling the dice.

Consideration of specific season and team IDs ($x_{ID,Season}$, $x_{ID,A}$, $x_{ID,B}$) as a feature does not significantly improve the metrics.

**Tab. 1:** Comparison of metrics for different models and features.

| Model | Features | Goal difference | | Total # of goals | |
|---|---|---|---|---|---|
| | | Cross entropy | RPS score | Cross entropy | RPS score |
| Poisson | $x^i_{\Delta G,AB}$ | 1.902±0.006 | - | 1.884±0.005 | - |
| Poisson | $x^i_{\Delta G,exp}, x^i_{\Sigma G,exp}$ | 1.901±0.006 | - | 1.884±0.005 | - |
| Neural network | $x^i_{\Delta G,exp}$ | 1.907±0.011 | 0.0445±0.0004 | 1.894±0.007 | 0.0614±0.0005 |
| Neural network | $x^i_{\Delta G,A}, x^i_{\Delta G,B}$ | 1.906±0.011 | 0.0444±0.0004 | 1.893±0.007 | 0.0613±0.0005 |
| Neural network | $x^i_{\Delta G,A}, x^i_{\Delta G,B}, x^i_{\Sigma G,A}, x^i_{\Sigma G,B}$ | 1.905±0.011 | 0.0444±0.0004 | 1.883±0.007 | 0.0606±0.0004 |
| Neural network | $x^i_{\Delta G,A}, x^i_{\Delta G,B}, x^i_{\Sigma G,A}, x^i_{\Sigma G,B}$, $x_{ID,A}, x_{ID,B}, x_{ID,Season}$ | 1.906±0.011 | 0.0444±0.0004 | 1.884±0.007 | 0.0606±0.0004 |
| Random Forest | $x^i_{\Delta G,exp}$ | 1.917±0.012 | 0.0445±0.0005 | 1.896±0.008 | 0.0611±0.0005 |
| Random Forest | $x^i_{\Delta G,A}, x^i_{\Delta G,A}$ | 1.916±0.010 | 0.0448±0.0004 | 1.891±0.007 | 0.0611±0.0005 |
| Random Forest | $x^i_{\Delta G,A}, x^i_{\Delta G,B}, x^i_{\Sigma G,A}, x^i_{\Sigma G,B}$ | 1.918±0.010 | 0.0445±0.0004 | 1.888±0.007 | 0.0608±0.0005 |
| Basic model | - | 2.08 | 0.0533 | 1.895 | 0.0616 |

### 4.7 Predicted distributions

The models output the goal difference or total number of goals with the highest probability as the final result. As goal differences with a value greater than 2 are largely due to random events, these are not predicted. This is shown in the confusion matrix in Fig. 1b.

Even though the metrics for the Poisson model and the neural network are very similar, there are still differences in the exact predictions of the models. To illustrate this, we look at $P(\Delta G)$ averaged over all games (Fig. 1c). In the Poisson model, draws ($\Delta G = 0$) are predicted with a lower probability than expected (Heuer 2012). Otherwise, the distribution is predicted very accurately. The neural network learns the distribution function directly from the data, so that $P(\Delta G)$ naturally matches the empirical distribution of the data set. For the total number of goals, both the prediction of the Poisson model and the neural network agree well with the empirical distribution of probabilities (Fig. 1d).

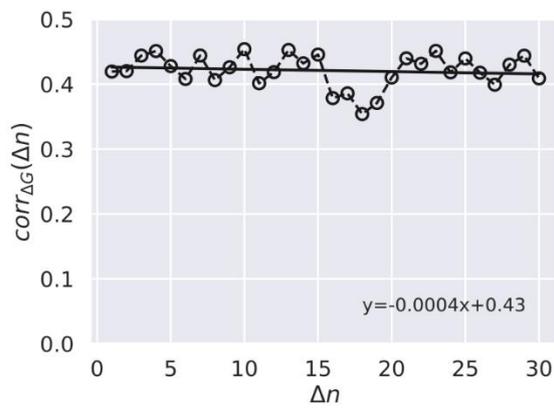
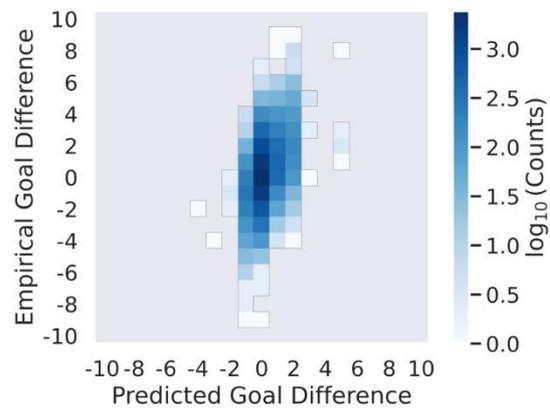
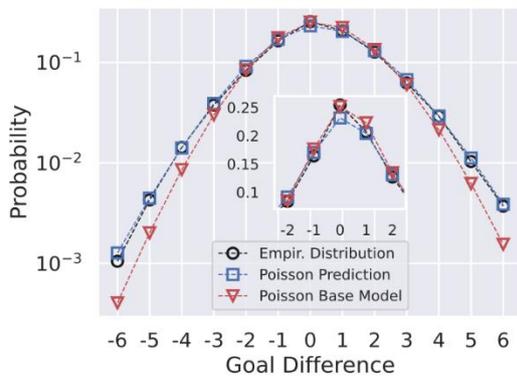
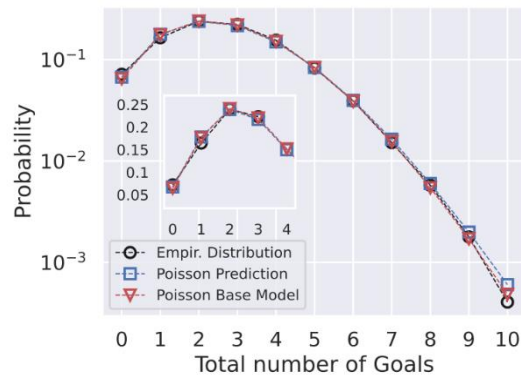

**Fig. 1:** a) Autocorrelation function of the goal differences. b) Confusion matrix for prediction and true distribution of the goal differences for the neural network model trained with the features $x^i_{+,A}, x^i_{-,A}, x^i_{+,B}, x^i_{-,B}$ for all 33 match days that are used for cross-validation. c) and d) Probability distribution of goal differences and total number of goals for the prediction of the Poison model and comparison with the Poisson base model. The inset shows a selected range on a linear scale.

## 5 Conclusion

This chapter illustrated how ML models can be used to predict the goal difference or total number of goals of a soccer match. Feature engineering was explained and the application of different features was discussed. The neural network is comparable to a Poisson model in terms of cross entropy and does not require any model assumptions. Compared to a base model, both models deliver significantly better results for the goal difference, while the results for the total number of goals are comparable to the base model. Conversely, the predictive power of goal chances as shown in (Heuer 2012, Heuer 2014) is significantly increased compared to the predictive power of goals. Therefore, the choice of model is less important than the quality of the data and the associated features in order to successfully predict match results.